\documentclass[aps,twocolumn,amsmath,amssymb,nopacs,prb,superscriptaddress,unsortedaddress]{revtex4}

\usepackage{epsf}
\usepackage{graphicx}
\usepackage{sidecap}

\def\3{2.8in}    
\def\2{2.5in}
\def\4{3.0in}

\def \beq {\begin{equation}}
\def \eeq {\end{equation}}
\begin{document}

\title{Observation of topological crystalline insulator phase in the lead tin chalcogenide Pb$_{1-x}$Sn$_x$Te material class}
\author{Su-Yang Xu}\affiliation {Joseph Henry Laboratory, Department of Physics, Princeton University, Princeton, New Jersey 08544, USA}
\author{Chang Liu}\affiliation {Joseph Henry Laboratory, Department of Physics, Princeton University, Princeton, New Jersey 08544, USA}
\author{N. Alidoust}\affiliation {Joseph Henry Laboratory, Department of Physics, Princeton University, Princeton, New Jersey 08544, USA}
\author{D. Qian}\affiliation {Joseph Henry Laboratory, Department of Physics, Princeton University, Princeton, New Jersey 08544, USA}\affiliation {Key Laboratory of Artificial Structures and Quantum Control (Ministry of Education), Department of Physics, Shanghai Jiao Tong University, Shanghai 200240, China.}
\author{M. Neupane}\affiliation {Joseph Henry Laboratory, Department of Physics, Princeton University, Princeton, New Jersey 08544, USA}

\author{J. D. Denlinger}\affiliation {Advanced Light Source, Lawrence Berkeley National Laboratory, Berkeley, California 94305, USA}

\author{Y. J. Wang}\affiliation {Department of Physics, Northeastern University, Boston, Massachusetts 02115, USA}
\author{H. Lin}\affiliation {Department of Physics, Northeastern University, Boston, Massachusetts 02115, USA}
\author{L. A. Wray}\affiliation {Joseph Henry Laboratory, Department of Physics, Princeton University, Princeton, New Jersey 08544, USA}\affiliation {Advanced Light Source, Lawrence Berkeley National Laboratory, Berkeley, California 94305, USA}
\author{R. J. Cava}\affiliation {Department of Chemistry, Princeton University, Princeton, New Jersey 08544, USA}
\author{A. Marcinkova} \affiliation{Department of Physics and Astronomy, Rice University, Houston, Texas 77005, USA}
\author{E. Morosan} \affiliation{Department of Physics and Astronomy, Rice University, Houston, Texas 77005, USA}
\author{A. Bansil}\affiliation {Department of Physics, Northeastern University, Boston, Massachusetts 02115, USA}
\author{M. Z. Hasan}\affiliation {Joseph Henry Laboratory, Department of Physics, Princeton University, Princeton, New Jersey 08544, USA}

\begin{abstract}
We perform systematic angle-resolved photoemission spectroscopic measurements on the lead tin telluride Pb$_{1-x}$Sn$_{x}$Te pseudobinary alloy system. We show that the (001) crystalline surface, which is a crystalline surface symmetric about the (110) mirror planes of the Pb$_{1-x}$Sn$_{x}$Te crystal, possesses four metallic surface states within its surface Brillouin zone. Our systematic Fermi surface and band topology measurements show that the observed Dirac-like surface states lie on the $\bar{\textrm{X}}-\bar{\Gamma}-\bar{\textrm{X}}$ momentum-space cuts. We further show that upon going to higher electron binding energies, the surface states' isoenergetic countours in close vicinity of each $\bar{\textrm{X}}$ point are observed to hybridize with each other, leading to a Fermi surface fractionalization due to a Lifshitz transition. In addition, systematic incident photon energy dependent measurements are performed, which enable us to unambiguously identify the surface states from the bulk bands. These systematic measurements of the surface and bulk electronic structure on Pb$_{1-x}$Sn$_{x}$Te, supported by our first principles calculation results, \textbf{\textit{for the first time}}, show that the Pb$_{1-x}$Sn$_{x}$Te system belongs to the topological crystalline insulator phase due to the four band inversions at the L points in its Brillouin zone, which has been recently theoretically predicted.
\end{abstract}
\date{\today}
\maketitle

The recent theoretical explorations on topological materials have extended the topological classification of insulating electronic band structures to a new topological phase --- topological crystalline insulators (TCIs) \cite{Liang PRL TCI}. A TCI phase is theoretically defined as a topologically nontrivial phase which arises from the crystal symmetries of the insulators. Historically, some form of spatial symmetry related topological property has already been observed in the Z$_2$ topological insulator Bi$_{1-x}$Sb$_x$ alloy system \cite{Mirror Chern Number, David Science BiSb}. However, such property in Bi$_{1-x}$Sb$_x$ system is completely masked by the dominating time-reversal symmetry based (Z$_2$) topological protection of surface states in the system. Up to this date, the role of crystal symmetries in topologically ordered phases remains experimentally elusive, especially the existence of topological order in insulators where Z$_2$ topological order is entirely absent. In this paper, we utilize angle-resolved photoemission spectroscopy (ARPES) to directly measure the surface state electronic structure of the lead tin telluride (Pb$_{1-x}$Sn$_{x}$Te) pseudobinary alloy system. Using the measured surface state electronic structure, we experimentally demonstrate the first realization of a topological crystalline insulator (TCI) phase, which is not a Z$_2$ (Kane-Mele) topological insulator.

\begin{figure*}
\centering
\includegraphics[width=14cm]{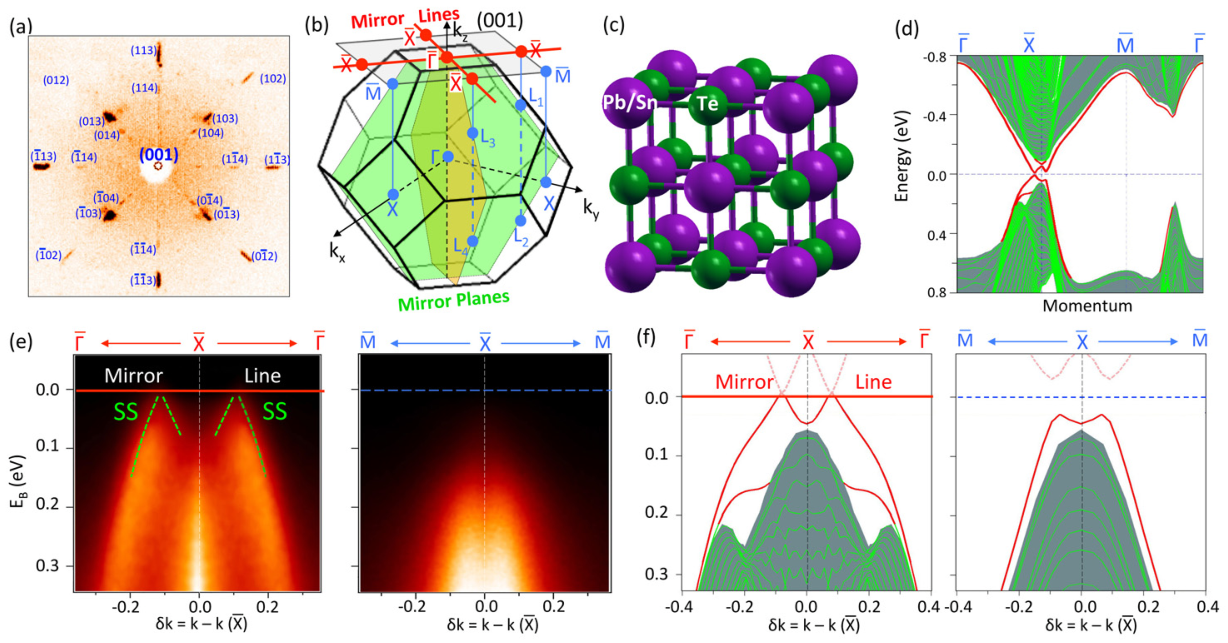}
\caption{\label{FCC} Mirror symmetry and surface states. (a) X-ray Laue diffraction pattern of a representative Pb$_{0.65}$Sn$_{0.35}$Te sample used for ARPES experiments. The Miller indices are noted for the diffraction peaks in Laue pattern, which reveals that the cleavage surface of the crystal is perpendicular to the [001] vertical crystal axis. (b) The first Brillouin zone (BZ) of Pb$_{0.65}$Sn$_{0.35}$Te lattice. The mirror planes are shown using green and light-brown colors. These mirror planes project onto the (001) crystal surface as the $\bar{\textrm{X}}-\bar{\Gamma}-\bar{\textrm{X}}$ mirror lines. (c) The Pb-rich Pb$_{1-x}$Sn$_{x}$Te crystalizes in the ideal ``sodium chloride'' crystal structure. (d) First principles based calculation of band dispersion of the stoichiometric end compound SnTe as a qualitative reference for the ARPES experiments. The surface states are shown by the red lines whereas the bulk band projections are represented by the shaded area. (e) High resolution ARPES dispersion maps measured in Pb$_{0.65}$Sn$_{0.35}$Te in the vicinity of $\bar{\textrm{X}}$ point along the $\bar{\Gamma}-\bar{\textrm{X}}-\bar{\Gamma}$ (the mirror line) and the $\bar{\textrm{M}}-\bar{\textrm{X}}-\bar{\textrm{M}}$ momentum space cuts respectively. The surface states cross Fermi level along the $\bar{\Gamma}-\bar{\textrm{X}}-\bar{\Gamma}$ cut (the mirror line) but lie well below along the $\bar{\textrm{M}}-\bar{\textrm{X}}-\bar{\textrm{M}}$ cut. (f) Calculation results of the stoichiometric SnTe end compound near $\bar{\textrm{X}}$ point. The chemical potential in the calculation is set roughly at the experimentally determined Fermi level as shown in (e).}
\end{figure*}

Lead tin telluride is a narrow band-gap semiconductor widely used for infrared optoelectronic and thermoelectric devices \cite{PbTe IR, PbTe Thermal}. Although this system has been under extensive studies in the past fifty years, neither any surface state nor any signature of topologically ordered character has been observed to the date \cite{PbTe IR, PbTe Thermal, PST band-gap, PST Inversion1, PST Inversion2, Volkov, Fradkin, Liang arXiv SnTe, SnTe ARPES, PST Rhombohedral Distortion}. Crystal structure of the Pb$_{1-x}$Sn$_{x}$Te system is based on the ``sodium chloride'' structure (space group Fm$\bar{3}$m (225)). In this structure, each of the two atom types (Pb/Sn, or Te) forms a separate face-centered cubic lattice, with the two lattices interpenetrating so as to form a three-dimensional checkerboard pattern as shown in Fig.~\ref{FCC}(c). The first Brillouin zone (BZ) of the crystal structure is a ``truncated octahedron'' with six square faces and eight hexagonal faces [Fig.~\ref{FCC}(b)]. The band-gap of Pb$_{1-x}$Sn$_{x}$Te is found to be a direct gap located at the L points in the BZ \cite{PST band-gap}. The L points are the centers of the eight hexagonal faces of the BZ. Due to the inversion symmetry of the crystal, each L point and its diametrically opposite partner on the BZ are completely equivalent. Thus there are four distinct L point momenta, as noted by L$_1$ through L$_4$ in Fig.~\ref{FCC}(b). It is known from both theoretical calculations and optical measurements that the band-gap at the four L points closes itself and re-opens upon increasing $x$ in the Pb$_{1-x}$Sn$_{x}$Te system \cite{PST Inversion1, PST Inversion2, Volkov, Fradkin}. The fact that band inversion occurs at even number of points within the BZ excludes the possibility of the Z$_2$-type (Kane-Mele) topological insulator phase in the Pb$_{1-x}$Sn$_{x}$Te system under ambient pressure. However, it is interesting to note that any two of the four L points along with the $\Gamma$ point form a momentum-space mirror plane within the BZ. For example, L$_1$, L$_2$ and $\Gamma$ point form the mirror plane (green) shown in Fig.~\ref{FCC}(b). The electronic states in this mirror plane are the eigenstates of the mirror operator $\mathcal{M}$ with respect to the symmetry plane. This fact provides a clue that the band inversion in the Pb$_{1-x}$Sn$_{x}$Te system may lead to a distinct topologically nontrivial phase that is irrelevant to the time-reversal operation $\mathcal{T}$ but may be the consequence of the spatial mirror symmetry operation $\mathcal{M}$. In that case, the crystal boundaries that are symmetric about the momentum-space mirror plane of Pb$_{1-x}$Sn$_{x}$Te may potentially host protected surface states within the inverted band-gap. In fact, a first principles electronic structure calculation study on the end product SnTe ($x=1.0$) suggested that the stoichiometric end product SnTe assumed in the ideal ``sodium chloride'' structure can host such nontrivial topological phase \cite{PST Inversion2, Volkov, Fradkin, Liang arXiv SnTe}. However, previous photoemisson studies have \textit{not} found any surface states on the surface of as-grown SnTe single crystals \cite{SnTe ARPES}. More importantly, the lattice of SnTe is reported to be subjected to a rhombohedral distortion \cite{PST Rhombohedral Distortion}, under which the mirror symmetries of the crystal are broken and the mirror protected TCI phase is not realized. We propose that the ideal system lies on the Pb-rich inverted side. We perform systematic surface and bulk electronic structure as well as crystalline symmetry studies on the lead tin telluride alloy system Pb$_{1-x}$Sn$_{x}$Te. Pb$_{0.65}$Sn$_{0.35}$Te is found to possess the ideal ``sodium chloride'' structure, which is also consistent with the previous structural studies \cite{PST Rhombohedral Distortion}. Since this composition is far away from the first principles calculation composition (SnTe), we carry out systematic angle-resolved photoemission studies on Pb$_{0.65}$Sn$_{0.35}$Te to systematically explore the band structure topology of the system. 

Fig.~\ref{FCC}(a) shows the X-ray Laue diffraction pattern of a representative Pb$_{0.65}$Sn$_{0.35}$Te sample used in our ARPES experiments. The four-fold symmetry of the Laue pattern reveals the ideal cubic structure of the crystal. The peaks in the Laue pattern are fitted based on a face-centered cubic lattice. The obtained Miller indices of the peaks clearly demonstrate that the ARPES measured cleavage surface is perpendicular to the [001] crystal direction. The (001) crystal surface, which is a surface symmetric about the momentum-space mirror planes, has a square surface BZ [the light-gray square in Fig.~\ref{FCC}(b)]. The $\Gamma$ and L points project onto the center of the square ($\bar{\Gamma}$) and the middle of the edges ($\bar{\textrm{X}}$), respectively. The mirror planes [green and light-brown planes in Fig.~\ref{FCC}(b)], critical for the topological order in the system, project onto the $\bar{\textrm{X}}-\bar{\Gamma}-\bar{\textrm{X}}$ mirror lines on the (001) surface BZ, as highlighted by the red lines in Fig.~\ref{FCC}(b). 

Since the low-energy physics of the system is dominated by the band inversion at L points (L points projected onto $\bar{\textrm{X}}$ points on the (001) surface), we start by showing the electronic structure measurements zoomed-in in the vicinity of one $\bar{\textrm{X}}$ point in Pb$_{0.65}$Sn$_{0.35}$Te, attempting to search for surface states within the band-gap. Fig.~\ref{FCC}(e) shows the ARPES dispersion maps near the $\bar{\textrm{X}}$ point along the $\bar{\Gamma}-\bar{\textrm{X}}-\bar{\Gamma}$ and the $\bar{\textrm{M}}-\bar{\textrm{X}}-\bar{\textrm{M}}$ momentum space cuts, respectively. The electronic structure along $\bar{\textrm{M}}-\bar{\textrm{X}}-\bar{\textrm{M}}$ is found to be fully gapped with the chemical potential lying within the band-gap. In sharp contrast, a pair of in-gap surface states are observed on the Fermi level along the $\bar{\Gamma}-\bar{\textrm{X}}-\bar{\Gamma}$ cut, which is the mirror line of the system. It is interesting to note that although the band inversion momenta (the L points) project exactly onto the $\bar{\textrm{X}}$ points, the observed surface states caused by the inversion are found to locate slightly away from the $\bar{\textrm{X}}$ point in momentum-space. As a qualitative guide to the ARPES experiments, we have also performed first principles electronic structure calculation on the stoichiometric end compound SnTe assuming ideal ``sodium chloride'' structure without rhombohedral distortion. Fig.~\ref{FCC}(f) presents the calculated electronic structure zoomed-in near the $\bar{\textrm{X}}$ point. The calculation results show qualitative correspondence with the ARPES measurements [Fig.~\ref{FCC}(e)]: a pair of surface states is found  on the Fermi level along the $\bar{\Gamma}-\bar{\textrm{X}}-\bar{\Gamma}$ mirror line direction, but absent along the $\bar{\textrm{M}}-\bar{\textrm{X}}-\bar{\textrm{M}}$ cut. 

\begin{figure*}
\centering
\includegraphics[width=14cm]{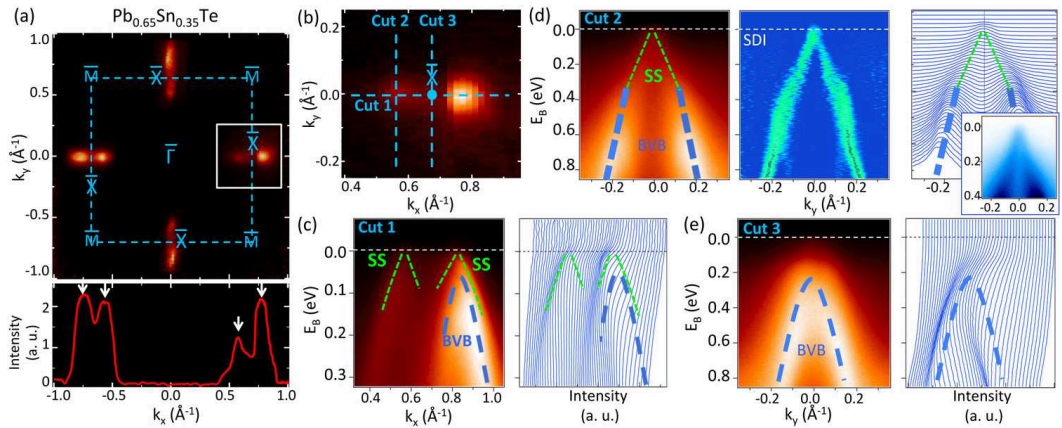}
\caption{\label{Basic_SS} (a) Top panel: Fermi surface mapping ($\textrm{E}_{\textrm{B}}=0.02$ eV) covering the first surface Brillouin zone (BZ). The surface states are only observed along the $\bar{\textrm{X}}-\bar{\Gamma}-\bar{\textrm{X}}$ cuts (the mirror lines), but are absent along all other high symmetry directions (e.g. $\bar{\textrm{M}}-\bar{\Gamma}-\bar{\textrm{M}}$ or $\bar{\textrm{M}}-\bar{\textrm{X}}-\bar{\textrm{M}}$). Four surface state Fermi pockets are enclosed within one surface BZ as observed in our measurements. Bottom panel: Spectral intensity distribution as a function of momentum along the horizontal mirror line (defined by $k_y=0$). (b) High resolution Fermi surface mapping ($\textrm{E}_{\textrm{B}}=0.02$ eV) in the vicinity of one of the $\bar{\textrm{X}}$ points [indicated by the white square in panel (a)]. (c)-(e) Dispersion maps ($\textrm{E}_{\textrm{B}}$ vs $k_{//}$) and corresponding energy (momentum) distribution curves of the momentum space cuts 1, 2, and 3. The second derivative image (SDI) of the measured dispersion is additionally shown for (d). The green and blue dotted lines are guides to the eye for the topological surface states (SS) and the bulk valence bands (BVB) respectively. Surface states are observed on the Fermi level for cuts 1 and 2. In particular, cut 2 (where the surface states are nearly isolated from the bulk valence band) reveals the nearly Dirac-like nature of the surface states up to the Fermi level. Inset of (d) (cut 2) shows the zoom-in of the dispersion map near the Fermi level, focusing on the Dirac-like surface states.}
\end{figure*}

We perform comprehensive measurements on the surface states of Pb$_{0.65}$Sn$_{0.35}$Te, and systematically map out the electronic structure. Fig.~\ref{Basic_SS}(a) shows the wide-range Fermi surface mapping ($\textrm{E}_{\textrm{B}}=0.02$ eV) covering the first surface BZ. The surface states are observed to be present, and only present, along the mirror line ($\bar{\Gamma}-\bar{\textrm{X}}-\bar{\Gamma}$) directions. No other states are found along any other momentum directions on the Fermi level. In close vicinity to each $\bar{\textrm{X}}$ point, a pair of surface states are observed along the mirror line direction. One lies inside the first surface BZ but the other is located outside. Therefore, in total four surface states are observed within the first surface BZ, which is in good agreement with four band inversions in the Pb$_{0.65}$Sn$_{0.35}$Te system. The Fermi surface mapping zoomed-in near the $\bar{\textrm{X}}$ point [Fig.~\ref{Basic_SS}(b)] reveals two unconnected small pockets (dots). The momentum space distance from the center of each pocket to the $\bar{\textrm{X}}$ point is about $0.09$ $\textrm{\AA}^{-1}$. Dispersion measurements ($\textrm{E}_{\textrm{B}}$ vs $k_{//}$) are performed along three different momentum space cuts, namely cuts 1, 2, and 3 defined in Fig.~\ref{Basic_SS}(b). In cut 1 [Fig.~\ref{Basic_SS}(c)], which is the mirror line ($\bar{\Gamma}-\bar{\textrm{X}}-\bar{\Gamma}$) direction, a pair of surface states are observed on the Fermi level. The bulk valence band is also present in cut 1 and is found to be located in close vicinity of the topological surface states. On the other hand, in cut 2 [Fig.~\ref{Basic_SS}(d)], the bulk band is pushed down to deep binding energies, which leaves the topological surface states close to the Fermi level nearly isolated. This enables us to study the detailed electronic dispersion of the surface states. Both the dispersion maps and the momentum distribution curves in cut 2 clearly reveal that the topological surface states are nearly Dirac-like (linearly dispersive). The linearly dispersive behavior is observed to preserve  from $\textrm{E}_{\textrm{B}}{\simeq}0.3$ eV all the way up to the Fermi level $\textrm{E}_{\textrm{B}}=0.0$ eV. Fitting of the momentum distribution curves of cut 2 shows that the experimental chemical potential ($\textrm{E}_\textrm{F}$) lies roughly at the Dirac point energy ($\textrm{E}_\textrm{D}$), $\textrm{E}_\textrm{F}=\textrm{E}_\textrm{D}\pm0.02$ eV. Assuming that the ARPES measured chemical potential does reflect the chemical potential deep inside the bulk (band bending effect is likely to be small due to the large dielectric constant of PbTe \cite{PbTe Dielectric}), then such near Dirac point Fermi level is highly favorable for future transport studies on these new topological surface states. At binding energy $\textrm{E}_{\textrm{B}}{\simeq}0.3$ eV, a sudden band velocity change is observed in cut 2, corresponding to the binding energy at which the topological surface states terminate and the bulk electronic bands start. The surface states' velocity is obtained to be $2.8{\pm}0.1$ $\textrm{eV}{\cdot}\textrm{\AA}$ ($4.2\pm0.2$ ${\times}10^5$ m/s) along cut 2, and $1.1{\pm}0.3$ $\textrm{eV}{\cdot}\textrm{\AA}$ ($1.7\pm0.4$ $\times10^5$ m/s) along cut 1, respectively.

\begin{figure*}
\centering
\includegraphics[width=14cm]{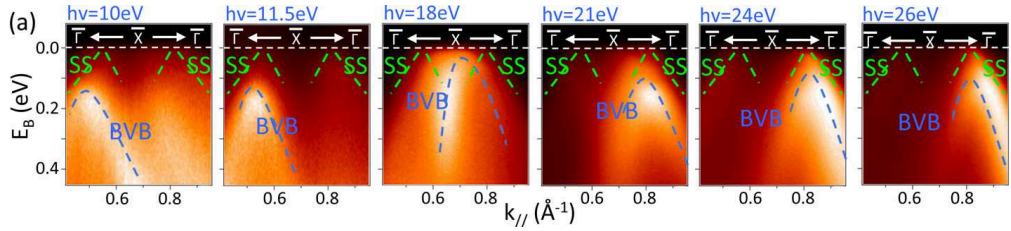}
\caption{\label{hv} Incident photon energy dependent measurements as a method for systematically isolating the surface states from the three-dimensional bulk bands. (a) ARPES dispersion maps as a function of the incident photon energy value. The green and blue dotted lines are guides to the eye for the topological surface states (SS) and the bulk valence bands (BVB) respectively. No significant dispersion is observed for the surface states near the Fermi level, whereas the bulk electronic bands are observed to show strong dispersion upon varying the incident photon energy.}
\end{figure*}

In order to systematically identify the surface states from the bulk electronic bands, we perform ARPES measurements as a function of incident photon energy. Upon varying the incident photon energy, one can effectively probe the electronic structure at different out-of-plane momentum $k_z$ values in a three-dimensional Brillouin zone. Through such approach, features in the ARPES measurements originating from bulk initial states can be distinguished from those originating from surface initial states since the bulk states are dispersive along the $k_z$ direction whereas the surface states are not. As shown in Fig.~\ref{hv}(a), a pair of surface states along the ($\bar{\Gamma}-\bar{\textrm{X}}-\bar{\Gamma}$) mirror line direction is observed at all photon energies. No observable dispersion of the surface states is found while changing the photon energy value. On the other hand, the bulk valence band is observed to show strong dispersion at different the photon energies (effectively different $k_z$ values). Both the binding energy value and the in-plane momentum location of the valence band maximum (VBM) are found to change significantly depending on the photon energy applied. These measurements enable us to unambiguously separate the surface states from the bulk bands. In addition, at photon energy of 18 eV, the in-plane momentum location of the VBM is observed to be very close to the $\bar{\textrm{X}}$ point, which is the (001) projection of the L point. Meanwhile, the binding energy value of the VBM at photon energy of 18 eV is found to be the closest to the Fermi level among all shown photon energies. This fact proves that the Pb$_{1-x}$Sn$_{x}$Te system has a direct band gap locating at the L points of its Brillouin zone.

\begin{figure*}
\centering
\includegraphics[width=14cm]{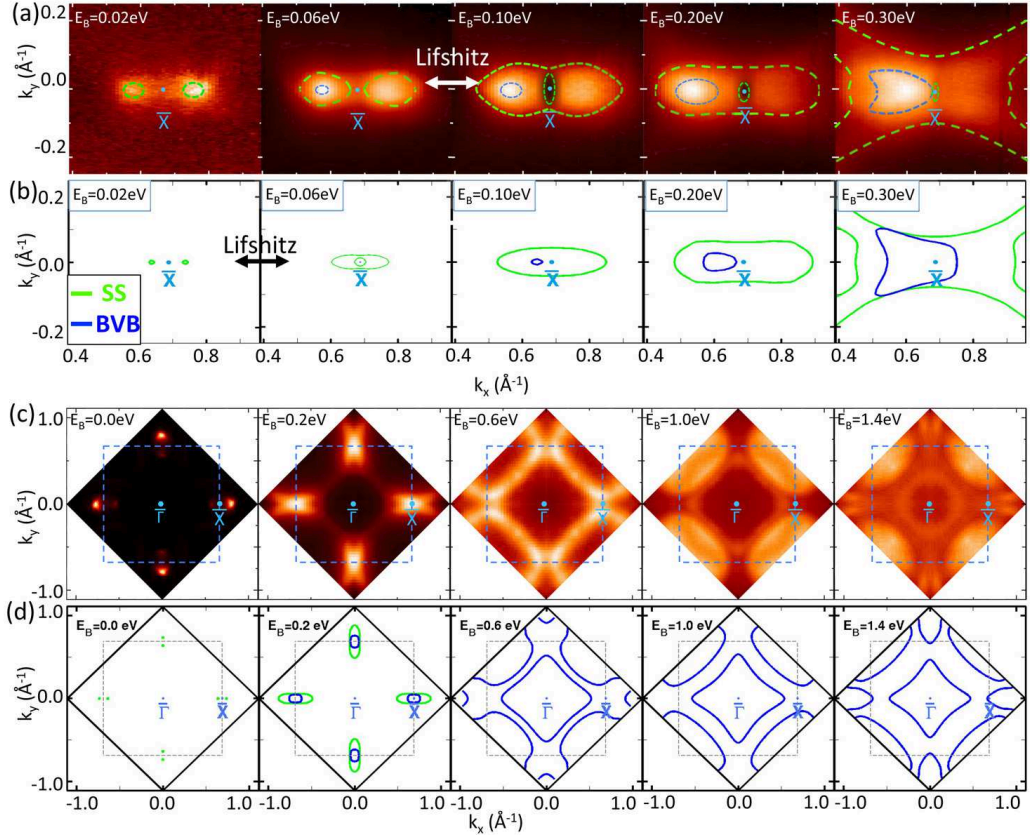}
\caption{\label{FS} Lifshitz transition on the surface. (a) and (b) Measured (a) and calculated (b) iso-energetic contour mappings as a function of electron binding energy in the vicinity of an $\bar{\textrm{X}}$ point. The green and blue dotted lines are guides to the eye for the topological surface states and the bulk bands respectively. The surface states' iso-energetic contours undergo a Lifshitz transition upon changing the binding energy: the topology of the iso-energetic contours changes from two unconnected contours (without enclosing $\bar{\textrm{X}}$) to two concentric contours both enclosing $\bar{\textrm{X}}$ point. The double-side-arrows in (a) and (b) indicate the Lifshitz transition, where the surface iso-energetic contour goes through a topological change. Calculations are performed on SnTe as a qualitative reference for comparison with the experimental data. The solid green and blue lines in calculation panels represent the surface states and the bulk bands respectively. (c) and (d) Measured (c) and calculated (d) iso-energetic contour mappings covering the first surface BZ as a function of binding energy.}
\end{figure*}

We present systematic measurements of our observed surface states' Fermi surface topology. Fig.~\ref{FS}(a) shows high-resolution iso-energetic contour mappings zoomed-in in the vicinity of an $\bar{\textrm{X}}$ point as a function of electron binding energy ($\textrm{E}_{\textrm{B}}$). Near the Fermi level ($\textrm{E}_{\textrm{B}}=0.02$ eV), two unconnected pockets are observed along the mirror line on opposite sides of the $\bar{\textrm{X}}$ point. In going to higher binding energies, the pockets grow larger and are observed to be about to touch each other at $\textrm{E}_{\textrm{B}}{\simeq}0.06$ eV. While further increasing the binding energy, naively the two pockets will overlap and cross each other. To avoid such crossing, the surface states' iso-energetic contours are observed to undergo a Lifshitz transition \cite{Lifshitz}: the topology is observed to change from two unconnected pockets to two concentric pockets both enclosing the $\bar{\textrm{X}}$ point. The transition is observed to take place in between binding energy $\textrm{E}_{\textrm{B}}=0.06$ eV to $\textrm{E}_{\textrm{B}}=0.10$ eV. While before the transition (0.0 eV $<\textrm{E}_{\textrm{B}}<$ 0.06 eV) the two unconnected pockets are observed to be hole-like, after the transition the inner one of the two concentric pockets is found to be electron-like (the size of the inner pocket is observed to decrease in going to higher binding energies). If one assumes that the Lifshitz transition does not exist in the system, then once the two hole-like pockets overlap, they will always cross each other upon going to higher binding energies. In that case, no electron-like contour should be present. Therefore the observation of the electron-like inner contour provides strong evidence that the Lifshitz transition indeed takes place in the system. The first principles calculated iso-energetic contours on SnTe (assuming no rhombohedral distortion) in Fig.~\ref{FS}(b) show qualitative agreement with the ARPES measurements. The bulk valence bands are also mapped out in a wide momentum space window as shown in Fig.~\ref{FS}(c). 

In summary, we have performed systematic surface and bulk electronic structure measurements on the (001) surface of the lead tin telluride Pb$_{0.65}$Sn$_{0.35}$Te system. We observe four surface states within its surface Brillouin zone, all of which are found to lie on the $\bar{\Gamma}-\bar{\textrm{X}}-\bar{\Gamma}$ momentum space mirror directions. The two-dimensional nature of the observed surface states is unambiguously identified by the incident photon energy dependent measurements. In addition, Fermi surface fractionalization (the Lifshitz transition) is observed on the surface states in close vicinity of each $\bar{\textrm{X}}$ point. These systematic surface states electronic structure measurements for the first time experimentally demonstrate that the Pb$_{1-x}$Sn$_{x}$Te pseudobinary alloy system is a topological crystalline insulator, which arises from the mirror symmetries of its ``sodium chloride''' crystal structure.

\bigskip

\newpage

\newpage

\newpage

\end{document}